\documentclass[12pt]{article}
\usepackage{latexsym}
\usepackage{amsmath,amsfonts}
\usepackage{times}
\allowdisplaybreaks[4]

\catcode`\@=11

\newcount\hour
\newcount\minute
\newtoks\amorpm \hour=\time\divide\hour by 60\minute
=\time{\multiply\hour by 60 \global\advance\minute by-\hour}
\edef\standardtime{{\ifnum\hour<12 \global\amorpm={am}%
        \else\global\amorpm={pm}\advance\hour by-12 \fi
        \ifnum\hour=0 \hour=12 \fi
        \number\hour:\ifnum\minute<10
        0\fi\number\minute\the\amorpm}}
\edef\militarytime{\number\hour:\ifnum\minute<10 0\fi\number\minute}

\def\draftlabel#1{{\@bsphack\if@filesw {\let\thepage\relax
   \xdef\@gtempa{\write\@auxout{\string
      \newlabel{#1}{{\@currentlabel}{\thepage}}}}}\@gtempa
   \if@nobreak \ifvmode\nobreak\fi\fi\fi\@esphack}
        \gdef\@eqnlabel{#1}}
\def\@eqnlabel{}
\def\@vacuum{}
\def\marginnote#1{}
\def\draftmarginnote#1{\marginpar{\raggedright\scriptsize\tt#1}}
\def\draft{
        \pagestyle{plain}
        \overfullrule=2pt
        \oddsidemargin -.5truein
        \def\@oddhead{\sl \phantom{\today\quad\militarytime} \hfil
        \smash{\Large\sl DRAFT} \hfil \today\quad\militarytime}
        \let\@evenhead\@oddhead
        \let\label=\draftlabel
        \let\marginnote=\draftmarginnote
        \def\ps@empty{\let\@mkboth\@gobbletwo
        \def\@oddfoot{\hfil \smash{\Large\sl DRAFT} \hfil}
        \let\@evenfoot\@oddhead}
        \def\@eqnnum{(\theequation)\rlap{\kern\marginparsep\tt\@eqnlabel}%
        \global\let\@eqnlabel\@vacuum}  }

\newcommand{\rf}[1]{(\ref{#1})}
\renewcommand{\theequation}{\thesection.\arabic{equation}}
\renewcommand{\thefootnote}{\fnsymbol{footnote}}
\newcommand{\newsection}{   
\setcounter{equation}{0}\section}

\def\appendix#1{\addtocounter{section}{1}\setcounter{equation}{0}
\renewcommand{\thesection}{\Alph{section}}
\section*{Appendix \thesection\protect\indent \parbox[t]{11.15cm}{#1}}
\addcontentsline{toc}{section}{Appendix \thesection\ \ \ #1}}

\def\be{\begin{equation}}
\def\ee{\end{equation}}
\def\beq{\begin{eqnarray}}
\def\eeq{\end{eqnarray}}

\def\parline{\,\partial\kern -0.55em /\,\,}

\def\half{{\frac{1}{2}}}

\def\CC{{\cal C}}

\def\LL{{\cal L}}

\def\NN{{\cal N}}

\def\abf{{\bf a}}
\def\bbf{{\bf b}}
\def\cbf{{\bf c}}
\def\dbf{{\bf d}}

\def\nubf{{\bf nu}}

\def\ibf{{\bf i}}
\def\iibf{{\bf ii}}
\def\iiibf{{\bf iii}}
\def\ivbf{{\bf iv}}
\def\vbf{{\bf v}}
\def\vibf{{\bf vi}}

\def\phik{|\phi\rangle}
\def\phibr{\langle\phi|}

\def\xik{|\xi\rangle}

\def\smL{{\scriptscriptstyle L}}

\def\smponetwo{{\scriptscriptstyle [1,2]}}

\def\sm(A)dS{{\scriptscriptstyle (A)dS }}

\def\Lb{\bar{L}}
\def\fb{\bar{f}}
\def\gb{\bar{g}}

\def\alphab{\bar\alpha}
\def\upsilonb{\bar\upsilon}

\def\eb{\bar{e}}

\def\irm{{\rm i}}

\def\diff{{\rm diff}}
\def\field{{\rm field}}

\def\mun{{\underline{m}}}

\begin{document}


\begin{flushright}
FIAN-TD-2019-04  \hspace{0.9cm} \ \ \ \ \  \\
arXiv: 1903.10495 [hep-th] \\
\end{flushright}

\vspace{1cm}

\begin{center}

{\Large \bf Light-cone continuous-spin field in AdS space}

\vspace{2.5cm}

R.R. Metsaev%
\footnote{ E-mail: metsaev@lpi.ru
}

\vspace{1cm}

{\it Department of Theoretical Physics, P.N. Lebedev Physical
Institute, \\ Leninsky prospect 53,  Moscow 119991, Russia }

\vspace{3.5cm}

{\bf Abstract}

\end{center}

We develop further the general light-cone gauge approach in AdS space and apply it for studying continuous-spin field. For such field, we find light-cone gauge Lagrangian and realization of relativistic symmetries. We find a simple realization of spin operators entering our approach. Generalization of our results to the gauge invariant Lagrangian formulation is also described. We conjecture that, in the framework of AdS/CFT,  the continuous-spin AdS field is dual to light-ray conformal operator. For some particular cases, our continuous-spin field leads to reducible models. We note two reducible models. The first model consists of massive scalar, massless vector, and partial continuous-spin field involving fields of all spins greater than one, while the second  model consists of massive vector, massless spin-2 field, and partial continuous-spin field involving all fields of spins greater than two.

\vspace{2cm}

Keywords: continuous-spin field; higher-spin field.

\newpage
\renewcommand{\thefootnote}{\arabic{footnote}}
\setcounter{footnote}{0}

\newsection{\large Introduction}

In view of the aesthetic features,  continuous-spin field has attracted some interest in recent time. For review, see Refs.\cite{Bekaert:2006py,Bekaert:2017khg,Brink:2002zx}. Extensive list of references on earlier studies of this theme may be found in Refs.\cite{Buchbinder:2019iwi,Buchbinder:2019esz}.
Alternative points of view on the role of continuous-spin field in string theory are presented in Refs.\cite{Savvidy:2003fx,Font:2013hia}. Interrelation of continuous-spin field and massive higher-spin field is discussed in Refs.\cite{Khan:2004nj}-\cite{Rehren:2017xzn}. Interacting continuous-spin fields are considered in Refs.\cite{Metsaev:2017cuz}-\cite{Metsaev:2018moa}, while various BRST Lagrangian formulations are studied in Refs.\cite{Bengtsson:2013vra,Metsaev:2018lth,Buchbinder:2018yoo}. Continuous-spin field in AdS space was investigated in Refs.\cite{Metsaev:2016lhs}-\cite{Metsaev:2017myp}. Other various important aspects of continuous-spin field were discussed in Refs.\cite{Schuster:2014hca}-\cite{Ponomarev:2010st}.

Continuous-spin field is decomposed into infinite chain of scalar, vector, and tensor fields which consists of every field just once. A similar infinite chain of fields appears in higher-spin gauge field theories in AdS space \cite{Vasiliev:1990en}. Other example of dynamical system involving infinite number of fields is a string theory. Light-cone gauge formulation simplifies considerably superstring action in AdS space \cite{Metsaev:2000yf,Metsaev:2000yu,Uvarov:2009hf}. We think that  light-cone gauge formulation will simplify study of continuous-spin field and therefore will be useful for better understanding of various aspects of continuous-spin field.

In this paper, we develop further our light-cone gauge formulation of AdS fields in Refs.\cite{Metsaev:1999ui,Metsaev:2003cu}. Namely, we obtain representation of the 4th-order Casimir operator of the $so(d,2)$ algebra in terms of spin operators entering a light-cone gauge Lagrangian. This allows us to express two constant parameters entering the light-cone gauge Lagrangian of continuous-spin field entirely in terms of the eigenvalues of the 2nd- and 4th-order Casimir operator of the $so(d,2)$ algebra. Such representation for the Lagrangian and a suitable parametrization of eigenvalues of the Casimir operators make the whole study more transparent and straightforward and considerably simplify analysis of classical unitarity and irreducibility of continuous-spin field. We obtain simple representation for spin operators entering our light-cone gauge approach. Also we make conjecture about holographic duality between continuous-spin bulk field and light-ray conformal operator.
Interrelations of light-cone gauge and gauge invariant approaches allow us to extend all our light-cone gauge results to the gauge invariant Lagrangian of continuous-spin field in a rather straightforward way. In due course, we discuss two models of continuous-spin fields which, besides partial continuous-spin fields, involve interesting spectrum of low-spin fields.

\newsection{ \large General light-cone gauge approach in $AdS_{d+1}$ space}

General light-cone gauge approach in AdS space was developed in Refs.\cite{Metsaev:1999ui,Metsaev:2003cu}. In this section, first, we review the formulation obtained in Ref.\cite{Metsaev:2003cu} and, second, we present our new result regarding the light-cone gauge representation for 4th-order Casimir operator of the $so(d,2)$ algebra.

Let $\phi(x,z)$ denote bosonic fields of arbitrary spin and type of symmetry. Collecting the fields into a ket-vector $|\phi\rangle$, we present a light-cone
gauge action in the following form:%
\footnote{ We use metric of $AdS_{d+1}$ space $ds^2=(-dx_0^2+dx_i^2+dx_{d-1}^2+dz^2)/z^2$. Light-cone coordinates $x^\pm$ are defined as $x^\pm=(x^{d-1} \pm x^0)/\sqrt{2}$, where
$x^+$ is a light-cone time. Our conventions for the coordinates and derivatives are as follows: $x^I= (x^i, x^d$), $x^d \equiv z$,
$\partial^i=\partial_i\equiv\partial/\partial x^i$,
$\partial_z\equiv\partial/\partial z$, $\partial^\pm=\partial_\mp
\equiv \partial/\partial x^\mp$, where we use the indices
$i,j =1,\ldots, d-2$; $I,J,K,L=1,2,\ldots,d-2, d$. Vectors of $so(d-1)$
algebra are decomposed as $X^I=(X^i,X^z)$, $X^I Y^I = X^i Y^i + X^zY^z$. Also we use the shortcuts, $x^Ix^I=x^ix^i+z^2$, $x^I\partial^I = x^i\partial^i+z\partial_z\,, \partial^I\partial^I=\partial^i\partial^i+\partial_z^2$.}
\be \label{16032019-manus-01}
S  = \int dz d^dx \langle \phi|\bigl(\Box + \partial_z^2 - \frac{1}{z^2}A\bigr)|\phi\rangle\,, \qquad \Box = 2\partial^+\partial^- + \partial^i\partial^i \,,
\ee
$\phibr = (\phik)^\dagger$, where operator $A$ being independent of space-time coordinates and their
derivatives is acting only on spin indices of $|\phi\rangle$. In general, fields entering  the $\phik$ are complex-valued.

The choice of the light-cone gauge spoils
the relativistic $so(d,2)$ symmetries of fields in $AdS_{d+1}$. Therefore in order to demonstrate that $so(d,2)$ symmetries are still present we find the
Noether charges which generate them. For free fields, Noether charges (generators)
have the following representation in terms of the $\phik$:
\be \label{18032019-manus-01}
G_\field = 2 \int dz dx^- d^{d-2} x\langle\partial^+\phi|G_\diff|\phi\rangle\,,
\ee
where $G_\diff$ stands for differential operators acting on $\phik$.  Action \rf{16032019-manus-01} is invariant under the transformations $\delta\phik =G_\diff\phik$. The  operators $G_\diff$ are given by
\beq
\label{18032019-manus-02} && P^i=\partial^i\,, \qquad  P^+=\partial^+\,, \hspace{1cm} P^-=-\frac{\partial^I\partial^I}{2\partial^+} +\frac{1}{2z^2\partial^+}A\,,
\\
\label{18032019-manus-03} && J^{+-} = x^+ P^- -x^-\partial^+\,, \hspace{1.3cm} J^{ij} = x^i\partial^j-x^j\partial^i + M^{ij}\,,
\\
\label{18032019-manus-04} && J^{+i}=x^+\partial^i-x^i\partial^+\,, \hspace{1.9cm} J^{-i}=x^-\partial^i-x^i P^- +M^{-i}\,,
\\
\label{18032019-manus-06}&& D = x^+ P^- +x^-\partial^++x^I\partial^I + \frac{d-1}{2}\,,
\\
\label{18032019-manus-07} && K^+ = -\frac{1}{2}(2x^+x^-+x^Jx^J)\partial^+ + x^+D\,,
\\
\label{18032019-manus-08} && K^i = -\frac{1}{2}(2x^+x^-+x^Jx^J)\partial^i +x^i D+M^{iJ}x^J+M^{i-}x^+\,,
\\
\label{18032019-manus-11} && K^-=-\frac{1}{2}(2x^+x^- + x^I x^I) P^- + x^-D+\frac{1}{\partial^+}x^I\partial^JM^{IJ}
-\frac{x^i}{2z\partial^+}[M^{zi},A] +\frac{1}{\partial^+}B\,,\qquad
\\
\label{18032019-manus-12} && \hspace{0.8cm} M^{-i} \equiv M^{iJ}\frac{\partial^J}{\partial^+}
-\frac{1}{2z\partial^+}[M^{zi},A]\,,\qquad M^{-i}=-M^{i-}\,,\quad M^{IJ}=- M^{JI}\,,
\eeq
where in \rf{18032019-manus-02}-\rf{18032019-manus-12} and below, we use operators $A$, $B$, $B^I$ $M^{IJ}$,  which, being independent of space-time coordinates and derivatives
are acting on spin indices of the ket-vector $\phik$. Our conventions for commutators of operators \rf{18032019-manus-02}-\rf{18032019-manus-12} may be found in relations (A6),(A7) in Ref.\cite{Metsaev:2015rda}. Note that we use the decompositions $M^{IJ} = M^{zi}, M^{ij}$, $B^I = B^z, B^i$. The operators $M^{IJ}$ and $B^I$ constitute a base of spin operators, while the operators $A$, $B$ can be expressed as
\beq
\label{18032019-manus-13} A & = & \CC_2 + 2B^z + 2M^{zi}M^{zi}+\frac{1}{2}M^{ij}M^{ij} +\frac{d^2-1}{4}\,,
\\
\label{18032019-manus-14} B & = & B^z + M^{zi}M^{zi}\,,
\eeq
where $\CC_2$ stands for an eigenvalue of the 2nd-order Casimir operator of the $so(d,2)$ algebra. The operators $B^I$, $M^{IJ}$ satisfy the commutators
\beq
\label{18032019-manus-15} && [M^{IJ}, M^{KL}] =\delta^{JK} M^{IL} + 3 \hbox{ terms}\,, \qquad [B^I,M^{JK}] =\delta^{IJ}B^K - \delta^{IK}B^J\,,
\\
\label{18032019-manus-16} && [B^I,B^J] = \Bigl( \CC_2 + \frac{1}{2} M^2 + \frac{d^2-3d+4}{2} \Bigr)M^{IJ} - (M^3)^{[I|J]}\,,
\eeq
where $\delta^{IJ}=\delta^{ij},\delta^{zz}$, $\delta^{zz}=1$ and we use the notation
\be \label{18032019-manus-17} M^2 \equiv M^{IJ} M^{IJ}\,, \qquad (M^3)^{[I|J]} \equiv \half M^{IK}M^{KL}M^{LJ}-  (I\leftrightarrow J)\,.
\ee
Note also that we use the following hermicity conjugation rules for the spin operators:

\vspace{-0.3cm}
\be \label{18032019-manus-18-a1}
M^{IJ\dagger} = - M^{IJ}\,, \qquad B^{I\dagger} = B^I\,.
\ee
From \rf{18032019-manus-15}, we see that the $M^{IJ}$ are spin operators of the $so(d-1)$ algebra, while the operator $B^I$ transforms as vector operator under transformations of the $so(d-1)$ algebra. It is the commutators \rf{18032019-manus-16} that are basic equations of light-cone gauge formulation of relativistic dynamics in $AdS_{d+1}$. We now ready to formulate our new result in this Section. We find that the 4th-order Casimir operator of the $so(d,2)$ algebra is expressed in terms of the spin operators $B^I$, $M^{IJ}$ as follows

\vspace{-0.4cm}
\be \label{18032019-manus-19}
\CC_4  =   B^IB^I - \half \bigl( \CC_2 + \frac{d^2 - 3d +4}{4} \bigr) M^2   - \frac{1}{8} (M^2)^2 - \frac{1}{4}  M^{IJ}M^{JK}M^{KL}M^{LI}\,,
\ee
where $M^2\equiv M^{IJ} M^{IJ}$. Our conventions for the Casimir operators are given in Appendix. Relations \rf{18032019-manus-13}-\rf{18032019-manus-19} are valid for bosonic and fermionic fields. For fermionic AdS fields, the discussion of general light-cone gauge action and operators $G_\diff$  may be found in Sec.4 in Ref.\cite{Metsaev:2003cu}.

\newsection{ \large Light-cone gauge continuous-spin AdS field}\label{lc-action}

To discuss bosonic light-cone gauge continuous-spin field we introduce a ket-vector%
\footnote{ For oscillators and vacuum, we use the rules: $[\alphab^I,\alpha^J] =\delta^{IJ}$, $[\upsilonb,\upsilon] =1$, $\alpha^{I\dagger} = \alphab^I$,  $\upsilon^\dagger = \upsilonb$, $\alphab^I |0\rangle = 0$, $\upsilonb |0\rangle = 0$.}

\vspace{-0.3cm}
\be \label{18032019-manus-20}
\phik =  \sum_{n=0}^\infty \frac{\upsilon^n}{ n!\sqrt{n!} } \alpha^{I_1}\ldots \alpha^{I_n} \phi^{I_1 \ldots I_n}(x,z) |0\rangle\,,
\ee
where, in \rf{18032019-manus-20}, fields with $n=0$, $n=1$, and $n\geq 2$ are the respective scalar, vector, and rank-$n$ totally symmetric traceless tensor fields of the $so(d-1)$ algebra. All fields in \rf{18032019-manus-20} are taken to be complex-valued. Ket-vector \rf{18032019-manus-20} satisfies the algebraic constraints

\vspace{-0.3cm}
\be
\label{18032019-manus-21} (N_\alpha - N_\upsilon) \phik =0 \,, \qquad \alphab^I\alphab^I\phik = 0\, \qquad N_\alpha \equiv \alpha^I\alphab^I\,, \qquad N_\upsilon \equiv \upsilon \upsilonb\,.
\ee
Our aim is to find realization of the spin operators $B^I$, $M^{IJ}$ on space of ket-vector $\phik$ \rf{18032019-manus-20}.
Realization of the $so(d-1)$ algebra spin operator $M^{IJ}$ on space of $\phik$ \rf{18032019-manus-20} is well known,

\vspace{-0.5cm}
\beq
\label{18032019-manus-23} && M^{IJ} = \alpha^I \alphab^J - \alpha^J\alphab^I\,,
\\
\label{18032019-manus-24} && (M^3)^{[I|J]}= \bigl( -\frac{1}{2}M^2+\frac{(d-3)(d-4)}{2} \bigr) M^{IJ} \,, \qquad M^2\equiv M^{IJ}M^{IJ}\,,
\eeq
where, in \rf{18032019-manus-24}, we show useful relation for $M^{IJ}$. Using \rf{18032019-manus-24}, we see that equations \rf{18032019-manus-16} and representation for $\CC_4$ \rf{18032019-manus-19} are considerably simplified as

\vspace{-0.5cm}
\beq
\label{18032019-manus-25} && [B^I,B^J] =  \bigl( \CC_2 + M^2 +2d - 4 \bigr) M^{IJ}\,,
\\
\label{18032019-manus-26} && \CC_4 = B^IB^I  -  \half ( \CC_2 + d -2 ) M^2 - \frac{1}{4} (M^2 )^2 \,.
\eeq
All that is required is to find solution to Eqs.\rf{18032019-manus-25},\rf{18032019-manus-26}. Solution to these equations is found to be

\vspace{-0.5cm}
\beq
\label{18032019-manus-27} && B^I = g \alphab^I + A^I \gb \,, \qquad  A^I \equiv \alpha^I - \alpha^J\alpha^J \frac{1}{2N_\alpha+d-1} \alphab^I\,,
\\
\label{18032019-manus-29} && g = g_\upsilon\upsilonb \,, \hspace{1.4cm} \gb = \upsilon \gb_\upsilon\,,
\\
\label{18032019-manus-30} && g_\upsilon = \NN_\upsilon f_\upsilon\,, \qquad  \gb_\upsilon = \NN_\upsilon \fb_\upsilon\,, \qquad \NN_\upsilon  = \bigl((N_\upsilon+1)(2N_\upsilon+d-1)\bigr)^{-1/2}\,,
\\
\label{18032019-manus-31} && f_\upsilon \fb_\upsilon = F_\upsilon\,,
\\
\label{18032019-manus-32} && F_\upsilon = F_0 - \bigl( \CC_2 +  d-1 \bigr)  N_\upsilon (N_\upsilon+ d-2) + N_\upsilon^2 (N_\upsilon+ d-2)^2\,,
\\
\label{18032019-manus-33} && F_0 = \CC_4\,,
\eeq
where $F_0$ \rf{18032019-manus-32} is a constant, while quantities $f_\upsilon$, $\fb_\upsilon$ \rf{18032019-manus-30},\rf{18032019-manus-31} depend on $N_\upsilon$ \rf{18032019-manus-21}.
We note that relations \rf{18032019-manus-27}-\rf{18032019-manus-32} are obtained from \rf{18032019-manus-25}, while the constant $F_0$ \rf{18032019-manus-33} is fixed by equation \rf{18032019-manus-26}.  From  \rf{18032019-manus-32},\rf{18032019-manus-33}, we see that our light-cone approach allows us to express the $F_\upsilon$ entirely in terms of eigenvalues of the Casimir operators. This fact turns out to be very important for our analysis of equation \rf{18032019-manus-31} and the condition for $B^I$ in \rf{18032019-manus-18-a1}. Relations  \rf{18032019-manus-27}-\rf{18032019-manus-33} exhaust all restrictions imposed on $f_\upsilon$, $\fb_\upsilon$ by equations \rf{18032019-manus-25},\rf{18032019-manus-26}. Remaining restrictions on $f_\upsilon$, $\fb_\upsilon$ can be obtained by using \rf{18032019-manus-31} and requirement of classical unitarity and irreducibility of continuous-spin field.

\noindent {\bf Irreducible classically unitary continuous-spin field}. Hermicity condition for $B^I$ \rf{18032019-manus-18-a1} leads to the relation $f_\upsilon^\dagger = \fb_\upsilon$. This relation and \rf{18032019-manus-31} imply that $\Im F_\upsilon(n)=0$, $F_\upsilon(n)\geq 0$ for all $n=0,1,\ldots,\infty$, where $F_\upsilon(n) \equiv F_\upsilon|_{N_\upsilon=n}$. Field \rf{18032019-manus-20} with such $F_\upsilon$ is referred to as classically unitary continuous-spin field.
Field \rf{18032019-manus-20}  with  $F_\upsilon$ \rf{18032019-manus-32} that satisfies the restrictions
\be \label{18032019-manus-34}
\Im F_\upsilon(n) = 0 \,, \qquad F_\upsilon(n) > 0 \,, \qquad \hbox{ for all } \ n=0,1,\ldots,\infty,
\ee
is referred to as irreducible classically unitary continuous-spin field. We use conventions in Ref.\cite{NIST}: $\Im F$ and $\Re F$ stand for the respective imaginary and real parts of $F$. To analyze \rf{18032019-manus-34} we use the Casimir operators (see \rf{17032019-manus-07},\rf{17032019-manus-08} in Appendix A)  and labels $p$, $q$ defined as

\vspace{-0.3cm}
\be \label{18032019-manus-35}
E_0 = \frac{d}{2} + p \,, \qquad s = \frac{2-d}{2} + q\,,
\ee
where the labels $p$, $q$ are complex-valued. It is the labels $p$, $q$ that considerably simplify our analysis of equations \rf{18032019-manus-34} and make our study transparent.  Plugging \rf{18032019-manus-35} into
\rf{17032019-manus-07},\rf{17032019-manus-08}, we find

\vspace{-0.4cm}
\beq
\label{18032019-manus-36} && \CC_2  = p^2 + q^2 - \frac{d^2+(d-2)^2}{4}\,,
\\
\label{18032019-manus-37} && \CC_4 = \bigl( p^2 - \frac{(d-2)^2}{4} \bigr) \bigl( q^2 - \frac{(d-2)^2}{4} \bigr)\,.
\eeq
In turn, plugging \rf{18032019-manus-36},\rf{18032019-manus-37} into \rf{18032019-manus-32}, we find the following factorized forms of $F_\upsilon$:

\vspace{-0.4cm}
\beq
F_\upsilon & = &  \bigl( (N_\upsilon+ \frac{d-2}{2})^2 - p^2\bigr) \bigl( (N_\upsilon+ \frac{d-2}{2})^2 - q^2\bigr)
\nonumber\\
\label{18032019-manus-38} & = & l_p l_{-p} l_q l_{-q}\,, \qquad\qquad l_X \equiv N_\upsilon + \frac{d-2}{2} + X\,.
\eeq
In view of $F_\upsilon(n) \equiv F_\upsilon|_{N_\upsilon=n}$, we note that the 1st equation in \rf{18032019-manus-34} amounts to the equations

\vspace{-0.3cm}
\be \label{18032019-manus-39}
\Im(p^2 + q^2) = 0 \,, \qquad \Im  (p^2q^2) = 0\,.
\ee
All non-trivial solutions to restrictions \rf{18032019-manus-39} are well-known,
\beq
&& \hspace{-1cm} \ibf: \ \ \Re p =0, \ \ \Re q=0; \hspace{1cm}  \iibf: \ \ p^*= q; \hspace{3cm} \iiibf: \ \ p^*= -q\,;
\nonumber\\[-11pt]
\label{18032019-manus-40} &&
\\[-9pt]
&& \hspace{-1cm} \ivbf: \ \ \Re p =0,\ \  \Im q=0; \hspace{0.8cm} \vbf: \ \ \Im p=0, \ \  \Re q=0\,; \hspace{1.1cm} \vibf: \ \ \Im p=0, \ \  \Im q=0.\qquad
\nonumber
\eeq
Using \rf{18032019-manus-38}, and $f_\upsilon^\dagger=\fb_\upsilon$, we see that all solutions to equation \rf{18032019-manus-31} corresponding to the respective cases in \rf{18032019-manus-40} can be presented as
\beq
\label{18032019-manus-42} && \hspace{-1.5cm} \ibf:  \hspace{1.3cm} f_\upsilon = l_p l_q\,, \hspace{2.4cm} \fb_\upsilon = l_{-p}l_{-q}\,, \hspace{2.5cm} \Re p =0\,,\quad \Re q=0;
\\
&&  \hspace{0.3cm} f_\upsilon = l_p l_{-p}\,, \hspace{2.1cm} \fb_\upsilon = l_{p^*}l_{-p^*}\,,
\nonumber\\[-10pt]
\label{18032019-manus-43} && \hspace{-1.5cm} \iibf,\ \iiibf: \hspace{9.4cm} p^*= \pm q;
\\[-10pt]
&&  \hspace{0.3cm} f_\upsilon = l_p l_{-p^*}\,, \hspace{2cm} \fb_\upsilon = l_{p^*}l_{-p}\,,
\nonumber\\
\label{18032019-manus-44} &&  \hspace{-1.5cm} \ivbf: \hspace{1cm}  f_\upsilon = l_p (l_ql_{-q})^{1/2}\,, \hspace{1.1cm} \fb_\upsilon =  l_{-p} (l_ql_{-q})^{1/2}\,,\hspace{1.4cm} \Re p=0\,,\quad \Im q=0;
\\
\label{18032019-manus-45} &&  \hspace{-1.5cm} \vbf: \hspace{1.2cm}  f_\upsilon = l_q (l_pl_{-p})^{1/2}\,, \hspace{1.1cm} \fb_\upsilon =  l_{-q} (l_pl_{-p})^{1/2}\,, \hspace{1.4cm} \Im p=0\,,\quad \Re q=0;
\\
\label{18032019-manus-46} &&  \hspace{-1.5cm} \vibf: \hspace{1cm}  f_\upsilon = (l_pl_{-p} l_ql_{-q})^{1/2}\,, \hspace{0.7cm}  \fb_\upsilon =  (l_pl_{-p} l_q l_{-q})^{1/2}\,,\hspace{1cm} \Im p =0\,,\quad  \Im q=0\,.
\eeq
The $f_\upsilon$, $\fb_\upsilon$ in \rf{18032019-manus-42}-\rf{18032019-manus-45} are complex-valued, while $f_\upsilon$, $\fb_\upsilon$ in \rf{18032019-manus-46} are real-valued. Therefore solutions in \rf{18032019-manus-42}-\rf{18032019-manus-45} are realized on complex-valued fields \rf{18032019-manus-20}, while solution \rf{18032019-manus-46} can be realized on real-valued fields \rf{18032019-manus-20}. Simple form of $f_\upsilon$, $\fb_\upsilon$ \rf{18032019-manus-42}-\rf{18032019-manus-45} is a new result in this Section. The $p$, $q$ satisfy restrictions \rf{18032019-manus-42}-\rf{18032019-manus-46}. For solutions $\ivbf$, $\vbf$, $\vibf$ \rf{18032019-manus-44}-\rf{18032019-manus-46}, equations \rf{18032019-manus-34} impose additional restrictions on $p$, $q$. We now analyse those additional restrictions in turn.

\noindent {\bf Statement 1}. Solutions  \rf{18032019-manus-42},\rf{18032019-manus-43} describe irreducible classically unitary continuous-spin fields.

\noindent {\bf Statement 2}. Solutions  \rf{18032019-manus-44}-\rf{18032019-manus-46} respect equations \rf{18032019-manus-34} provided the $p$, $q$, besides restrictions in \rf{18032019-manus-44}-\rf{18032019-manus-46}, satisfy the following additional restrictions:
\beq
\label{18032019-manus-47} && \hspace{-1.5cm} \ivbf\!: \qquad  q^2 < x_0\,;
\\
\label{18032019-manus-48} && \hspace{-1.5cm} \vbf\!:\qquad p^2 < x_0\,;
\\
\label{18032019-manus-49} && \hspace{-1.5cm} \vibf\hbox{-}\abf\!:\quad  p^2 < x_0\,, \hspace{2.2cm} q^2 < x_0\,, \hspace{2.1cm} p^2 \ne q^2\,;
\\
\label{18032019-manus-50} && \hspace{-1.5cm}  \vibf\hbox{-}\bbf\!: \hspace{0.3cm}  x_n  < p_n^2 < x_{n+1}\,, \qquad x_n < q_n^2 < x_{n+1}\,, \hspace{0.7cm} p_n^2\ne q_n^2\,, \quad  n = 0,1,\ldots, \infty\,;\qquad
\\
\label{18032019-manus-51} && \hspace{-1.5cm}  \vibf\hbox{-}\cbf\!: \quad  p^2 \ne x_n \,; \hspace{2.2cm} p^2 = q^2\,, \hspace{4.1cm}  n = 0,1,\ldots,\infty\,;
\\
\label{18032019-manus-52} &&  x_n \equiv \bigl( n+\frac{d-2}{2} \bigr)^2\,.
\eeq
We see that, for solution in \rf{18032019-manus-46},  there are three classes of additional restrictions \rf{18032019-manus-49}-\rf{18032019-manus-51} on the labels $p$, $q$. The Statements can easily be proved by using \rf{18032019-manus-31},\rf{18032019-manus-38}.
Relations  \rf{18032019-manus-42}-\rf{18032019-manus-52} provide the complete description of  all irreducible classically unitary continuous-spin fields. As reducible case provides interesting field content we now proceed with the discussion of reducible classically unitary continuous-spin fields.

\noindent {\bf Reducible classically unitary continuous-spin field}. Continuous-spin field with  $F_\upsilon$ \rf{18032019-manus-38} that satisfies the equations
\beq
\label{18032019-manus-53} && \hspace{-1.4cm}  F_\upsilon(n_r) = 0  \hspace{4cm}  \hbox{ for some } n_r \in  0,1,\ldots, \infty\,,
\\
\label{18032019-manus-54} && \hspace{-1.4cm} \Im F_\upsilon(n) = 0 \,, \qquad F_\upsilon(n) > 0  \hspace{0.9cm} \hbox{ for all } n =0,1,\ldots, \infty \hbox{ and } n \ne n_r \hspace{1cm}
\eeq
is referred to as reducible classically unitary continuous-spin field.

\noindent {\bf Statement 3}. Solutions \rf{18032019-manus-44}-\rf{18032019-manus-46} respect equations \rf{18032019-manus-53},\rf{18032019-manus-54} provided the $p$, $q$, besides restrictions in \rf{18032019-manus-44}-\rf{18032019-manus-46}, satisfy the following additional restrictions
\beq
\label{18032019-manus-55-4} && \hspace{-2cm} \ivbf\hbox{-}1\!:\quad  q^2 = x_0\,;
\\
\label{18032019-manus-55-5} && \hspace{-2cm} \vbf\hbox{-}1\!:\hspace{0.7cm} p^2 = x_0\,;
\\
\label{18032019-manus-55-a01} && \hspace{-2cm} \vibf\hbox{-}1\abf\!:\quad  p^2 = x_0\,, \hspace{1.2cm}   q^2 < x_1\,;
\\
\label{18032019-manus-55-a02} && \hspace{-2cm} \vibf\hbox{-}1\bbf\!:\quad  q^2 = x_0\,, \hspace{1.2cm}  p^2 < x_1\,;
\\
\label{18032019-manus-55-a1} && \hspace{-2cm} \vibf\hbox{-}1\cbf\!:\quad  p_n^2 = x_n\,, \hspace{1.2cm} x_{n-1} < q_n^2 < x_{n+1}\,, \hspace{1.1cm}   n = 1,\ldots,\infty\,;
\\
\label{18032019-manus-55-a2} && \hspace{-2cm} \vibf\hbox{-}1\dbf\!:\quad  q_n^2 = x_n\,, \hspace{1.2cm} x_{n-1} < p_n^2 < x_{n+1}\,, \hspace{1.1cm}     n = 1,\ldots,\infty\,;
\\
\label{18032019-manus-55-a3} && \hspace{-2cm} \vibf\hbox{-}2\abf\!:\quad  p_n^2 = x_n\,, \hspace{1.2cm} q_n^2 = x_{n+1}\,, \hspace{2.5cm}  n = 0,1,\ldots,\infty\,;
\\
\label{18032019-manus-55-a4} && \hspace{-2cm}  \vibf\hbox{-}2\bbf\!: \quad  p_n^2 = x_{n+1}\,, \hspace{0.9cm} q_n^2 = x_n\,, \hspace{2.8cm} n = 0,1,\ldots, \infty\,;\qquad
\eeq
where $x_n$ is given in \rf{18032019-manus-52}. Relations \rf{18032019-manus-55-4}-\rf{18032019-manus-55-a2} are associated with one root of $F_\upsilon(n)$ \rf{18032019-manus-53}, while relations \rf{18032019-manus-55-a3},\rf{18032019-manus-55-a4} are associated with two roots of  $F_\upsilon(n)$ \rf{18032019-manus-53}.
The $f_\upsilon$, $\fb_\upsilon$ in \rf{18032019-manus-44}-\rf{18032019-manus-46} with the additional restrictions in \rf{18032019-manus-55-4}-\rf{18032019-manus-55-a4} describe decoupled fields. Namely, decomposing ket-vector $\phik$ \rf{18032019-manus-20} as
\beq
\label{18032019-manus-58-a0} \phik & = &  |\phi^{0,0}\rangle  + |\phi^{1,\infty}\rangle\,,  \hspace{3.4cm} \hbox{ for } \ \ivbf\hbox{-}1\,, \ \vbf\hbox{-}1\,, \  \vibf\hbox{-}1\abf\, \  \vibf\hbox{-}1\bbf;
\\
\label{18032019-manus-58} \phik & = &  |\phi^{0,n}\rangle  + |\phi^{n+1,\infty}\rangle\,,  \hspace{3cm} \hbox{ for } \ \vibf\hbox{-}1\cbf\,, \  \vibf\hbox{-}1\dbf;
\\
\label{18032019-manus-59} \phik & = &  |\phi^{0,n}\rangle  + |\phi^{n+1,n+1}\rangle + |\phi^{n+2,\infty}\rangle\,, \hspace{0.8cm} \hbox{ for } \ \vibf\hbox{-}2\abf\,, \ \vibf\hbox{-}2\bbf;
\\
\label{18032019-manus-60} && |\phi^{M,N}\rangle \equiv  \sum_{n=M}^N \frac{\upsilon^n}{n!\sqrt{n!}} \alpha^{I_1} \ldots \alpha^{I_n} \phi^{I_1\ldots I_n}(x,z) |0\rangle\,,
\eeq
we can verify that action \rf{16032019-manus-01} is decomposed into direct sum of actions for fields appearing on r.h.s in \rf{18032019-manus-58-a0}-\rf{18032019-manus-59}. Field $|\phi^{0,n}\rangle$ in \rf{18032019-manus-58-a0}-\rf{18032019-manus-59} is a spin-$n$ massive field, while field $|\phi^{n+1,n+1}\rangle$ in \rf{18032019-manus-59} is a spin-$(n+1)$ massless field. The eigenvalue of $\CC_2$ for these fields is given by $\CC_2 = p^2 + q^2 - x_0 - x_1$.  Mass square of the spin-$n$ field is given by $m^2 = p^2 + q^2 - x_n - x_{n-1}$ when $n>0$ and $m^2= \CC_2$ when $n=0$. Relations  \rf{18032019-manus-44}-\rf{18032019-manus-46} and \rf{18032019-manus-55-4}-\rf{18032019-manus-55-a4} provide the complete description of all reducible classically unitary continuous-spin fields.

\noindent {\bf Classically non-unitary reducible continuous-spin field and classically unitary irreducible partial continuous-spin field}.
In \rf{18032019-manus-58-a0}-\rf{18032019-manus-59}, we see appearance of shortened fields $|\phi^{{k+1},\infty}\rangle$. We refer to field $|\phi^{k,\infty}\rangle$ \rf{18032019-manus-60} with $k>0$ as depth-$k$ partial continuous-spin field. In \rf{18032019-manus-58-a0}-\rf{18032019-manus-59}, classically unitary irreducible partial continuous-spin fields enter reducible classically unitary field $\phik$ \rf{18032019-manus-20}. We note however that classically unitary irreducible partial continuous-spin fields may enter a reducible classically non-unitary field $\phik$. It is easy to understand that such cases can be obtained by considering the equations $\Im F_\upsilon(n) = 0$ for all $n=0,1,\ldots,\infty$ and
\be \label{18032019-manus-61}
F_\upsilon(k-1) <  0 \,,\quad F_\upsilon(k) = 0 \,,\quad F_\upsilon(n) > 0 \,, \quad \ n=k+1,k+2,\ldots,\infty,
\ee
$k\geq0$. The following Statement can easily be proved by using \rf{18032019-manus-31} and \rf{18032019-manus-38}.

\noindent {\bf Statement 4}. Solutions  \rf{18032019-manus-44}-\rf{18032019-manus-46} respect equations \rf{18032019-manus-61} provided the $p$, $q$, besides restrictions in \rf{18032019-manus-44}-\rf{18032019-manus-46}, satisfy the following additional restrictions:
\beq
\label{18032019-manus-62} && \hspace{-2cm} \ivbf\hbox{-}\nubf\hbox{-}1\!: \hspace{0.7cm}  q_k^2 = x_k\,, \hspace{3.9cm}   k = 1,\ldots,\infty\,;
\\
\label{18032019-manus-63} && \hspace{-2cm}  \vbf\hbox{-}\nubf\hbox{-}1\!: \hspace{0.8cm}  p_k^2 = x_k\,, \hspace{3.9cm} k = 1,\ldots, \infty\,;\qquad
\\
\label{18032019-manus-64} && \hspace{-2cm} \vibf\hbox{-}\nubf\hbox{-}2\abf\!:\quad  p_n^2 = x_n\,, \hspace{1.2cm} q_k^2 = x_k\,, \hspace{1.1cm}   k-n>1,\qquad n,k = 0,1,\ldots,\infty\,;
\\
\label{18032019-manus-65} && \hspace{-2cm} \vibf\hbox{-}\nubf\hbox{-}2\bbf\!:\quad  p_k^2 = x_k\,, \hspace{1.2cm} q_n^2 = x_n\,, \hspace{1.1cm} k-n > 1  \qquad    n,k = 0,1,\ldots,\infty\,;
\eeq
where $x_n$ is given in \rf{18032019-manus-52}. Expressions for $f_\upsilon$, $\fb_\upsilon$ in \rf{18032019-manus-44}-\rf{18032019-manus-46} with the additional restrictions in \rf{18032019-manus-62}-\rf{18032019-manus-65} describe the following decoupled fields:
\beq
\label{18032019-manus-66} \phik & = &  |\phi^{0,k}\rangle  + |\phi^{k+1,\infty}\rangle\,,  \hspace{3.7cm} \hbox{ for } \ \ivbf\hbox{-}\nubf\hbox{-}1\,, \  \vbf\hbox{-}\nubf\hbox{-}1;
\\
\label{18032019-manus-67} \phik & = &  |\phi^{0,n}\rangle  + |\phi^{n+1,k}\rangle + |\phi^{k+1,\infty}\rangle\,, \hspace{1.8cm} \hbox{ for } \ \vibf\hbox{-}\nubf\hbox{-}2\abf\,, \ \vibf\hbox{-}\nubf\hbox{-}2\bbf;
\eeq
where $|\phi^{0,k}\rangle$ \rf{18032019-manus-66} and $|\phi^{0,n}\rangle$ \rf{18032019-manus-67} are the respective classically non-unitary spin-$k$ and unitary spin-$n$ massive fields, $|\phi^{n+1,k}\rangle$ \rf{18032019-manus-67} is a spin-$k$ classically non-unitary partial-massless field, while  $|\phi^{k+1,\infty}\rangle$ \rf{18032019-manus-66},\rf{18032019-manus-67} is the irreducible classically unitary  partial continuous-spin field. Mass square of $|\phi^{n+1,k}\rangle$ \rf{18032019-manus-67} is given by $m^2 = x_n - x_{k-1}$.

The following remarks are in order.

\noindent \abf) For solution \rf{18032019-manus-42}, there are no restrictions on $\Im p$, $\Im q$ \rf{18032019-manus-42}. We note that the same happens for light-ray conformal operator. Namely, such operator is realized as unitary representation of the $so(d,2)$ algebra and  labelled by conformal dimension $\Delta = E_0$ and continuous-spin $s$ given in \rf{18032019-manus-35} with $\Re p=0$, $\Re q = 0$ and no restrictions on $\Im p$, $\Im q$. We recall that our solution \rf{18032019-manus-42} describes irreducible classically unitary continuous-spin field. {\it We then conjecture that, in the framework of AdS/CFT correspondence, our continuous-spin AdS field with $E_0$ and $s$ in \rf{18032019-manus-35} and solution \rf{18032019-manus-42} is dual to light-ray conformal operator having $\Delta=E_0$ and $s$ as in \rf{18032019-manus-35} with $\Re p=0$, $\Re q = 0$}. Also we expect that continuous-spin AdS fields associated with the solutions \rf{18032019-manus-43}-\rf{18032019-manus-46} are also dual to the respective conformal operators having conformal dimension $\Delta=E_0$ and $s$ as in \rf{18032019-manus-35}. Interesting discussion of light-ray operators may be found in Ref.\cite{Balitsky:2015tca}.

\noindent \bbf) Using \rf{18032019-manus-59},\rf{18032019-manus-60} for $n=0$, $n=1$, we note two models of reducible classically unitary continuous-spin field with interesting spectrum for low-spin finite-component fields,
\beq
\label{18032019-manus-67-a1}&& \phik   =    |\phi^{0,0}\rangle_{massive\, \atop scalar} + |\phi^{1,1}\rangle_{massless\atop  vector} + |\phi^{2,\infty}\rangle_{partial\atop contin-spin}\,,
\\
\label{18032019-manus-67-a2} && \phik   =    |\phi^{0,1}\rangle_{massive\atop  vector}  + |\phi^{2,2}\rangle_{massless\atop  spin-2} + |\phi^{3,\infty}\rangle_{partial\atop contin-spin}\,.
\eeq
Scalar field $|\phi^{0,0}\rangle$ \rf{18032019-manus-67-a1} has mass square $m^2=0$, $m^2 \ne m_c^2$, where $m_c^2=(1-d^2)/4$ is the mass square of conformal invariant scalar field. Therefore we refer to scalar field $|\phi^{0,0}\rangle$ \rf{18032019-manus-67-a1} as a massive field. Massive vector field  $|\phi^{0,1}\rangle$ \rf{18032019-manus-67-a2} has the mass square $m^2 = 2d$.

\noindent \cbf) By changing phases of complex-valued fields in \rf{18032019-manus-20}, the complex-valued $f_\upsilon$, $\fb_\upsilon$ \rf{18032019-manus-42}-\rf{18032019-manus-45} can be cast into real-valued form. The real-valued form of $f_\upsilon$, $\fb_\upsilon$ \rf{18032019-manus-42}-\rf{18032019-manus-45} is presented as in \rf{18032019-manus-46}, where $p$,$q$ are given in \rf{18032019-manus-42}-\rf{18032019-manus-45}. It is the real-valued form of $f_\upsilon$, $\fb_\upsilon$ \rf{18032019-manus-42}-\rf{18032019-manus-46} that we used
in Ref.\cite{Metsaev:2016lhs}. Our new complex-valued form of $f_\upsilon$, $\fb_\upsilon$ \rf{18032019-manus-42}-\rf{18032019-manus-45} is simpler that the one in \rf{18032019-manus-46}. For example $f_\upsilon$, $\fb_\upsilon$ \rf{18032019-manus-42} are degree-2 polynomials in  $N_\upsilon$ \rf{18032019-manus-21}, while $f_\upsilon$, $\fb_\upsilon$ \rf{18032019-manus-46} involve a square root of degree-4 polynomials in the $N_\upsilon$. Therefore it seems preferable to study the continuous-spin field by using $f_\upsilon$, $\fb_\upsilon$ given in \rf{18032019-manus-42}-\rf{18032019-manus-45}. It is the use of complex-valued fields in \rf{18032019-manus-20} that allows us to introduce simple representations for solutions in \rf{18032019-manus-42}-\rf{18032019-manus-45}.

\noindent {\bf Flat space}. We use the chance to discuss new representation for operators $f_\upsilon$, $\fb_\upsilon$ entering con\-ti\-nuous-spin fields in flat space. Light-cone gauge formulation of massless and massive continuous-spin fields in flat space $R^{d-1,1}$, $d$-arbitrary, was obtained in Refs.\cite{Metsaev:2017cuz,Metsaev:2018moa} . In the flat space  $R^{d-1,1}$, equation for $f_\upsilon$, $\fb_\upsilon$ \rf{18032019-manus-31} takes the form
\beq
\label{18032019-manus-53-a1} f_\upsilon \fb_\upsilon  = F_\upsilon \,, \qquad F_\upsilon = \kappa^2 - m^2 N_\upsilon( N_\upsilon + d-3)\,, \quad \kappa^2 > 0\,, \qquad m^2 \leq 0\,,
\eeq
$f_\upsilon^\dagger=\fb_\upsilon$, where the mass parameter $m$ and the continuous-spin parameter $\kappa$ are related to 2nd- and 4th-order Casimir operators of the Poincar\'e algebra (for details see Ref.\cite{Metsaev:2018moa}). Restrictions on the $m$ and $\kappa$ \rf{18032019-manus-53-a1} are obtained by requiring the classical unitarity and irreducibility. In Ref.\cite{Metsaev:2018moa}, we discussed the solution $f_\upsilon =\sqrt{F_\upsilon}$, $\fb_\upsilon =\sqrt{F_\upsilon}$. Such solution is realized on space of real-valued continuous-spin field. We now note the new solution given by
\beq
&& f_\upsilon =  \sqrt{-m^2} \bigl( N_\upsilon + \frac{d-3}{2} \bigr) + \irm \sigma_\kappa\,,  \qquad \fb_\upsilon =  \sqrt{-m^2} (N_\upsilon + \frac{d-3}{2}) - \irm \sigma_\kappa\,,  \qquad
\nonumber\\
\label{18032019-manus-54-a1} && \sigma_\kappa \equiv \bigl( \kappa^2 + \frac{(d-3)^2}{4}m^2 \bigr)^{1/2}\,,  \hspace{1.7cm} \hbox{ for } \ \ \sigma_\kappa^2 \geq 0 \,.
\eeq
The new solution \rf{18032019-manus-54-a1} is realized on space of complex-valued continuous-spin field. Our new solution enters the spin operator of massive continuous-spin field as follows. In expression for $g$ and $\gb$ in (2.21) in Ref.\cite{Metsaev:2018moa}, we  make the respective substitutions $\sqrt{F_\upsilon}\rightarrow f_\upsilon$ and $\sqrt{F_\upsilon}\rightarrow \fb_\upsilon$.

\newsection{ \large Gauge invariant action for continuous-spin AdS field}

Gauge invariant action for continuous-spin field was obtained in Ref.\cite{Metsaev:2016lhs}.
The gauge invariant action depends on two parameters. In this section, our aim is twofold. First, motivated by our result for light-cone gauge continuous-spin field we are going to express the two parameters in terms of the eigenvalues of two Casimir operators. Second, we discuss extension of our new representation for solutions given in \rf{18032019-manus-42}-\rf{18032019-manus-45} to the gauge invariant Lagrangian formulation.

Gauge invariant action is formulated in terms of ket-vector given by
\be \label{19032019-manus-01}
\phik = \sum_{n=0}^\infty \frac{\upsilon^n}{n!\sqrt{n!}} \alpha^{a_1} \ldots \alpha^{a_n} \phi^{a_1\ldots a_n}(x) |0\rangle\,,
\ee
where, in \rf{19032019-manus-01}, fields with $n=0$, $n=1$, and $n\geq 2$ are the respective scalar, vector, and rank-$n$ totally symmetric tensor fields the Lorentz algebra $so(d,1)$.   Fields with $n \geq 4$ are double-traceless, $\phi^{aabba_5\ldots a_n}=0$. All fields in \rf{19032019-manus-01} are taken to be complex-valued. In this Section and in Ref.\cite{Metsaev:2016lhs}, vector indices of the Lorentz algebra $so(d,1)$ take values $a,b=0,1,\ldots,d$.

Gauge invariant action and Lagrangian of continuous-spin field we found can be presented as%
\footnote{ Our conventions are: $e=\det e_\mun^a$, where $e_\mun^a$ is vielbein in AdS space; $\Box_{AdS}$ is the D'Alembert operator in AdS space; $\alpha D \equiv \alpha^a D^a$, $\alphab D \equiv \alphab^a D^a$, $\alpha^2 \equiv \alpha^a \alpha^a$, $\alphab^2 \equiv \alphab^a \alphab^a$, $N_\alpha \equiv \alpha^a\alphab^a$, where $D^a = e^{a \mun} D_\mun$ and $D_\mun$ is a covariant derivative in AdS. For more details of our notation, see Appendix A in Ref.\cite{Metsaev:2016lhs}.
}
\beq
\label{19032019-manus-02} S  & = &  \int d^{d+1}x\,\LL\,,  \qquad  \LL  =  e \phibr   E \phik \,,
\\
\label{19032019-manus-03} E & \equiv & (1-\frac{1}{4}\alpha^2 \bar\alpha^2) (\Box_{AdS} + m_1 + m_2
\alpha^2\bar\alpha^2) - L \Lb\,,
\\
\label{19032019-manus-04} && \Lb \equiv  \bar\alpha D - \half \alpha D  \bar\alpha^2  -
\eb_\smL\Pi^\smponetwo + \half e_\smL \bar\alpha^2\,,
\\
\label{19032019-manus-05} && L \equiv \alpha D  - \half \alpha^2 \bar\alpha D  - e_\smL \Pi^\smponetwo +
\half \eb_\smL \alpha^2\,,
\\
&& \hspace{1cm} \Pi^\smponetwo \equiv 1 - \alpha^2 \frac{1}{2(2N_\alpha + d + 1)}\alphab^2\,,
\eeq
where $\langle\phi| \equiv (\phik)^\dagger$, while quantities $m_1$, $m_2$, $e_\smL$, and $\eb_\smL$ are defined by the relations
\beq
\label{19032019-manus-06} && m_1  = - \mu_0 +   N_\upsilon(N_\upsilon+d-1) + 2d - 4  \,, \qquad m_2  =  -1 \,,
\\
\label{19032019-manus-07} && e_\smL =  \NN_\upsilon f_\upsilon \upsilonb\,, \qquad  \eb_\smL = - \upsilon \NN_\upsilon \fb_\upsilon\,,\hspace{2cm}\NN_\upsilon  \equiv \bigl((N_\upsilon+1)(2N_\upsilon+d-1)\bigr)^{-1/2},\qquad
\\
\label{19032019-manus-08} && f_\upsilon \fb_\upsilon  =  F_\upsilon\,, \hspace{1.1cm} f_\upsilon^\dagger = \fb_\upsilon\,,
\\
\label{19032019-manus-09} && F_\upsilon \equiv   \mu_1 -  (\mu_0 - d+3)  N_\upsilon (N_\upsilon + d-2) + N_\upsilon^2 (N_\upsilon +d-2)^2  \,,
\eeq
where $\mu_0$, $\mu_1$ stand for constant parameters. The quantity $F_\upsilon$ in \rf{19032019-manus-09} is the same as the one in light-cone gauge approach in \rf{18032019-manus-32}. Comparing \rf{19032019-manus-09} and \rf{18032019-manus-32},\rf{18032019-manus-33}, we can entirely express the parameters $\mu_0$ and $\mu_1$ in terms of the eigenvalues of the 2nd and 4th-order Casimir operators
\be \label{19032019-manus-10}
\mu_0 = \CC_2 + 2d -4\,, \qquad \mu_1 = \CC_4\,.
\ee
Using \rf{18032019-manus-36},\rf{18032019-manus-37}, and \rf{19032019-manus-10}, we can represent $F_\upsilon$ \rf{19032019-manus-09} as in \rf{18032019-manus-38}. This implies that  our whole analysis we carried out for classically (ir)reducible light-cone gauge continuous-spin field in Section \ref{lc-action} is automatically extended to the gauge invariant formulation in this Section. For example, all expressions  for $f_\upsilon$, $\fb_\upsilon$ entering  \rf{19032019-manus-06},\rf{19032019-manus-07} can be read from  \rf{18032019-manus-42}-\rf{18032019-manus-60}. Solutions \rf{18032019-manus-42}-\rf{18032019-manus-45} are realized on complex-valued fields \rf{19032019-manus-01}, while solution \rf{18032019-manus-46}
can be realized on real-valued field in \rf{19032019-manus-01}. Note that, in Ref.\cite{Metsaev:2016lhs}, we discuss solutions in the form given in
\rf{18032019-manus-46}.

Gauge symmetries are described by using ket-vector of gauge transformation parameters,
\be \label{19032019-manus-11}
\xik = \sum_{n=0}^\infty \frac{\upsilon^{n+1}}{n!\sqrt{(n+1)!}} \alpha^{a_1} \ldots \alpha^{a_n} \xi^{a_1\ldots a_n}(x) |0\rangle\,.
\ee
In \rf{19032019-manus-11}, gauge parameters with $n=0$, $n=1$, and $n\geq 2$ are the
respective scalar, vector, and totally symmetric rank-$n$ tensor fields of the Lorentz algebra $so(d,1)$, $\xi^{aaa_3\ldots a_n}=0$. Using ket-vectors $\phik$ and $\xik$ and operators $e_\smL$, $\eb_\smL$  \rf{19032019-manus-07}, we write  gauge transformations as
\be \label{19032019-manus-12}
\delta \phik  = G\xik\,, \qquad G = \alpha D - e_\smL - \alpha^2\frac{1}{2N_\alpha
+ d- 1}\eb_\smL \,.
\ee

\medskip

{\bf Conclusions}. In this paper, we developed further the general light-cone formulation in Ref.\cite{Metsaev:1999ui,Metsaev:2003cu} and applied it for studying free continuous-spin field in AdS space. Extension of our approach to interacting continuous-spin AdS field could be of great interest. In this respect, we note that, using the method in Ref.\cite{Metsaev:2005ar}, we studied interacting vertices of light-cone gauge continuous-spin field in flat space in Refs.\cite{Metsaev:2017cuz,Metsaev:2018moa}, while, in Ref.\cite{Metsaev:2018xip}, we developed the method for studying finite-component light-cone gauge AdS fields. We believe therefore that the results in this paper and the ones in Refs.\cite{Metsaev:2017cuz,Metsaev:2018moa,Metsaev:2018xip} will be helpful for studying interacting continuous-spin AdS field. For the reader's convenience, we note that
various BRST methods for studying interacting finite-component fields may be found in Refs.\cite{Fotopoulos:2008ka,Metsaev:2012uy}. Other interesting methods for investigation of interacting finite-component fields were developed in Refs.\cite{Manvelyan:2010jr,Vasilev:2011xf}. Study of continuous-spin field along the line of group-theoretical methods in Refs.\cite{Basile:2017kaz} could also be of some interest.

\bigskip
{\bf Acknowledgments}. Author would like to thank A. Tseytlin for reading the manuscript and
useful comments. This work was supported by the RFBR Grant No.17-02-00317.

\vspace{-0.3cm}
\setcounter{section}{0} \setcounter{subsection}{0}

\appendix{ \large Casimir operators of the $so(d,2)$ algebra}

In this Appendix, we explain our conventions for Casimir operators of the $so(d,2)$ algebra. To this end we start with a manifestly $(d+2)$-dimensional covariant approach. Generators of the $so(d,2)$ algebra denoted by $J^{AB}$ satisfy the commutators
\be \label{17032019-manus-01}
[J^{AB},J^{CE}]= \eta^{BC}J^{AE} + 3 \hbox{ terms},\qquad J^{AB\dagger} = -J^{AB}\,, \qquad \eta^{AB} =  (--,+,\ldots, +)\,,
\ee
where vector indices of the $so(d,2)$ algebra take values $A,B,C,E=0',0,1,2,\ldots,d$. In terms of the $J^{AB}$, the 2nd- and 4th-order Casimir operators of the $so(d,2)$ algebra are defined to be
\be \label{17032019-manus-02}
\CC_2 = \half J^{AB} J^{BA}\,,\qquad \CC_4 = \half \CC_2^2 + \frac{d(d-1)}{4} \CC_2 - \frac{1}{4}J^{AB} J^{BC} J^{CE} J^{EA}\,.
\ee

In Ref.\cite{Metsaev:1999ui}, we shown that the first relation in \rf{17032019-manus-02} allows us to find the relation for  the operator $A$ in \rf{18032019-manus-13}. We now note that by using the light-cone gauge generators of the $so(d,2)$ algebra given in relations  \rf{18032019-manus-02}-\rf{18032019-manus-14}, we verified that the 4th-order Casimir operator $\CC_4$ \rf{17032019-manus-02} takes the form given in \rf{18032019-manus-19}. For doing so, we should relate generators in \rf{17032019-manus-01} with the light-cone generators \rf{18032019-manus-02}-\rf{18032019-manus-12}. To this end we decompose $(d+2)$ coordinates $x^A$ as
\be  \label{17032019-manus-03}
x^A = x^\oplus,\ x^\ominus\,, \ x^a\,, \quad a=0,1,2,\ldots,d-1\,, \qquad   x^\oplus  \equiv \frac{1}{\sqrt{2}}(x^d +    x^{0^\prime})\,, \quad x^\ominus \equiv \frac{1}{\sqrt{2}}(x^d  -    x^{0^\prime})\,.
\ee
In the frame of the coordinates $x^{\oplus,\ominus}$, $x^a$,  the $J^{AB}$ and $\eta^{AB}$ \rf{17032019-manus-01} are represented as
\be \label{17032019-manus-04}
J^{AB} = J^{\oplus a}, \  J^{\ominus a}, \ J^{\ominus \oplus},\  J^{ab}\,, \qquad \eta^{AB} = \eta^{\oplus\ominus}, \eta^{\ominus\oplus}, \eta^{ab}\,,
\ee
where $\eta^{\oplus\ominus}=1$, $\eta^{\ominus\oplus}=1$. Now, using notation of the conformal algebra considered in the base of the algebra $so(d-1,1)$ spanned by $J^{ab}$, we identify generators \rf{17032019-manus-04} as:
\be  \label{17032019-manus-06}
P^a = J^{\oplus a}, \qquad  K^a =  J^{\ominus a}, \qquad D =  J^{\ominus \oplus}\,.
\ee
Generators \rf{17032019-manus-06} and $J^{ab}$ satisfy the commutators given in  (A6),(A7) in Ref.\cite{Metsaev:2015rda}, while relation (A8) in Ref.\cite{Metsaev:2015rda} provides the light-cone gauge decomposition of the generators \rf{17032019-manus-06}. Now using \rf{17032019-manus-02} and relations \rf{18032019-manus-02}-\rf{18032019-manus-12}, we verified that $\CC_4$ in \rf{17032019-manus-02} amounts to $\CC_4$ in \rf{18032019-manus-19}.

To parametrize eigenvalues of the Casimir operators for totally symmetric representations of the $so(d,2)$ algebra we use labels $E_0$, $s$. In conformal algebra notation, $E_0\equiv\Delta$, where $\Delta$ is conformal dimension of a conformal operator, while $s$ is associated with spin. In general, the $E_0$ and $s$ are complex-valued. Eigenvalues of the Casimir operators \rf{17032019-manus-02} are given by
\beq
\label{17032019-manus-07} \CC_2 & = &   E_0(E_0 - d) +  s(s+d-2)\,,
\\
\label{17032019-manus-08} \CC_4 & = &   (E_0 - 1)(E_0 - d + 1) s(s+d-2)\,.
\eeq


\small
\begin{thebibliography}{30}

\parskip=-6pt


\bibitem{Bekaert:2006py}
  X.~Bekaert and N.~Boulanger,
 ``The Unitary representations of the Poincare group in any spacetime dimension,''
   in 2nd Modave Summer School in Theoretical Physics Modave, Belgium, August 6-12, 2006, 2006.
  hep-th/0611263.


\bibitem{Bekaert:2017khg}
  X.~Bekaert and E.~D.~Skvortsov,
  Int.J.Mod.Phys.\ A {\bf 32}, no.23n24, 1730019 (2017)
  [arXiv:1708.01030].


\bibitem{Brink:2002zx}
  L.~Brink, A.~M.~Khan, P.~Ramond and X.~z.~Xiong,
  J.\ Math.\ Phys.\  {\bf 43}, 6279 (2002)
  [hep-th/0205145].

\bibitem{Buchbinder:2019iwi}
  I.~L.~Buchbinder, S.~Fedoruk and A.~P.~Isaev,
  ``Twistorial and space-time descriptions of massless infinite spin (super)particles and fields,''
  arXiv:1903.07947 [hep-th].

\bibitem{Buchbinder:2019esz}
  I.~L.~Buchbinder, S.~J.~Gates and K.~Koutrolikos,
  ``Superfield continuous spin equations of motion,''
  arXiv:1903.08631 [hep-th].



\bibitem{Savvidy:2003fx}
  G.~K.~Savvidy,
  Int.\ J.\ Mod.\ Phys.\ A {\bf 19}, 3171 (2004)
  [hep-th/0310085].
%
\\
%
  J.~Mourad,
  ``Continuous spin particles from a string theory,''
  hep-th/0504118.


\bibitem{Font:2013hia}
  A.~Font, F.~Quevedo and S.~Theisen,
  Fortsch.\ Phys.\  {\bf 62}, 975 (2014)
  [arXiv:1302.4771 [hep-th]].


\bibitem{Khan:2004nj}
  A.~M.~Khan and P.~Ramond,
  J.\ Math.\ Phys.\  {\bf 46}, 053515 (2005)
  Erratum: [J.\ Math.\ Phys.\  {\bf 46}, 079901 (2005)]
  [hep-th/0410107].


\bibitem{Bekaert:2005in}
  X.~Bekaert and J.~Mourad,
  JHEP {\bf 0601}, 115 (2006)
  [hep-th/0509092].


\bibitem{Rehren:2017xzn}
  K.~H.~Rehren,
  JHEP {\bf 1711}, 130 (2017)
  doi:10.1007/JHEP11(2017)130
  [arXiv:1709.04858 [hep-th]].

\bibitem{Metsaev:2017cuz}
  R.~R.~Metsaev,
  JHEP {\bf 1711}, 197 (2017)
  [arXiv:1709.08596 [hep-th]].


\bibitem{Bekaert:2017xin}
  X.~Bekaert, J.~Mourad and M.~Najafizadeh,
  JHEP {\bf 1711}, 113 (2017)
  [arXiv:1710.05788 [hep-th]].


\bibitem{Rivelles:2018tpt}
  V.~O.~Rivelles,
  ``A Gauge Field Theory for Continuous Spin Tachyons,''
  arXiv:1807.01812 [hep-th].

\bibitem{Metsaev:2018moa}
  R.~R.~Metsaev,
  JHEP {\bf 1812}, 055 (2018)
  [arXiv:1809.09075 [hep-th]].


\bibitem{Bengtsson:2013vra}
  A.~K.~H.~Bengtsson,
  JHEP {\bf 1310}, 108 (2013)
  [arXiv:1303.3799 [hep-th]].


\bibitem{Metsaev:2018lth}
  R.~R.~Metsaev,
  Phys.\ Lett.\ B {\bf 781}, 568 (2018)
  [arXiv:1803.08421 [hep-th]].


\bibitem{Buchbinder:2018yoo}
  I.L.Buchbinder, V.Krykhtin and H.Takata,
  Phys.\ Lett.\ B {\bf 785}, 315 (2018),
  arXiv:1806.01640 [hep-th].



\bibitem{Metsaev:2016lhs}
  R.~R.~Metsaev,
  Phys.\ Lett.\ B {\bf 767}, 458 (2017)
  [arXiv:1610.00657 [hep-th]].


\bibitem{Metsaev:2017ytk}
  R.~R.~Metsaev,
  Phys.\ Lett.\ B {\bf 773}, 135 (2017)
  [arXiv:1703.05780 [hep-th]].


\bibitem{Zinoviev:2017rnj}
  Y.~M.~Zinoviev,
  Universe {\bf 3}, no. 3, 63 (2017)
  [arXiv:1707.08832 [hep-th]].


\bibitem{Khabarov:2017lth}
  M.~V.~Khabarov and Y.~M.~Zinoviev,
  Nucl.\ Phys.\ B {\bf 928}, 182 (2018)
  [arXiv:1711.08223 [hep-th]].


\bibitem{Metsaev:2017myp}
  R.~R.~Metsaev,
  J.\ Phys.\ A {\bf 51}, no. 21, 215401 (2018)
  [arXiv:1711.11007 [hep-th]].


\bibitem{Schuster:2014hca}
  P.~Schuster and N.~Toro,
  Phys.\ Rev.\ D {\bf 91}, 025023 (2015)
  [arXiv:1404.0675 [hep-th]].


\bibitem{Najafizadeh:2015uxa}
  X.Bekaert, M.Najafizadeh, M.R.Setare,
  Phys.\ Lett.\ B {\bf 760}, 320 (2016)
  [arXiv:1506.00973 [hep-th]].

\bibitem{Rivelles:2014fsa}
  V.~O.~Rivelles,
  Phys.\ Rev.\ D {\bf 91}, no. 12, 125035 (2015)
  [arXiv:1408.3576 [hep-th]].
%
\\
%
  V.~O.~Rivelles,
  Eur.\ Phys.\ J.\ C {\bf 77}, no. 7, 433 (2017)
  [arXiv:1607.01316 [hep-th]].


\bibitem{Najafizadeh:2017tin}
  M.~Najafizadeh,
  Phys.\ Rev.\ D {\bf 97}, no. 6, 065009 (2018)
  [arXiv:1708.00827 [hep-th]].


\bibitem{Alkalaev:2017hvj}
  K.~B.~Alkalaev and M.~A.~Grigoriev,
  JHEP {\bf 1803}, 030 (2018)
  [arXiv:1712.02317 [hep-th]].


\bibitem{Buchbinder:2018soq}
  I.L.Buchbinder, S.Fedoruk, A.P.Isaev, A.Rusnak,
  JHEP {\bf 1807}, 031 (2018),
  arXiv:1805.09706 [hep-th]


\bibitem{Alkalaev:2018bqe}
  K.~Alkalaev, A.~Chekmenev and M.~Grigoriev,
  JHEP {\bf 1811}, 050 (2018)
  [arXiv:1808.09385 [hep-th]].



\bibitem{Ponomarev:2010st}
  D.~S.~Ponomarev and M.~A.~Vasiliev,
  Nucl.\ Phys.\ B {\bf 839}, 466 (2010)
  [arXiv:1001.0062 [hep-th]].



\bibitem{Vasiliev:1990en}
  M.~A.~Vasiliev,
  Phys.\ Lett.\ B {\bf 243}, 378 (1990).
%
\\
%
  M.~A.~Vasiliev,
  Phys.\ Lett.\  B {\bf 567}, 139 (2003)
  [arXiv:hep-th/0304049].




\bibitem{Metsaev:2000yf}
  R.~R.~Metsaev and A.~A.~Tseytlin,
  Phys.\ Rev.\ D {\bf 63}, 046002 (2001)
  [hep-th/0007036].



\bibitem{Metsaev:2000yu}
  R.~R.~Metsaev, C.~B.~Thorn and A.~A.~Tseytlin,
  Nucl.\ Phys.\ B {\bf 596}, 151 (2001)
  [hep-th/0009171].


\bibitem{Uvarov:2009hf}
  D.~V.~Uvarov,
  Nucl.\ Phys.\ B {\bf 826}, 294 (2010)
  [arXiv:0906.4699 [hep-th]].
%
\\
  D.~V.~Uvarov,
  Mod.\ Phys.\ Lett.\ A {\bf 25}, 1251 (2010)
  [arXiv:0912.1044 [hep-th]].


\bibitem{Metsaev:1999ui}
  R.~R.~Metsaev,
  Nucl.\ Phys.\ B {\bf 563}, 295 (1999)
  [hep-th/9906217].

\bibitem{Metsaev:2003cu}
  R.~R.~Metsaev,
  Phys.\ Lett.\ B {\bf 590}, 95 (2004)
  [hep-th/0312297].


\bibitem{Metsaev:2015rda}
  R.~R.~Metsaev,
  JHEP {\bf 1510}, 110 (2015)
  [arXiv:1507.06584 [hep-th]].



\bibitem{NIST}
Olver F. W. J. (ed.). NIST handbook of mathematical functions. Cambridge University Press, 2010


\bibitem{Balitsky:2015tca}
  I.Balitsky, V.Kazakov and E.Sobko,
  Phys. Rev. D {\bf 93}, no. 6, 061701 (2016)
  [arXiv:1506.02038].




\bibitem{Metsaev:2005ar}
  R.~R.~Metsaev,
  Nucl.\ Phys.\ B {\bf 759}, 147 (2006)
  [hep-th/0512342].



\bibitem{Metsaev:2018xip}
  R.~R.~Metsaev,
  Nucl.\ Phys.\ B {\bf 936}, 320 (2018)
  [arXiv:1807.07542 [hep-th]].



\bibitem{Fotopoulos:2008ka}
  A.~Fotopoulos and M.~Tsulaia,
  Int.\ J.\ Mod.\ Phys.\ A {\bf 24}, 1 (2009)
  [arXiv:0805.1346 [hep-th]].
%
\\
%
  P.~Dempster and M.~Tsulaia,
  Nucl.\ Phys.\ B {\bf 865}, 353 (2012)
  [arXiv:1203.5597 [hep-th]].
%
\\
%
  M.~Henneaux, G.~Lucena G\'omez and R.~Rahman,
  JHEP {\bf 1208}, 093 (2012)
  [arXiv:1206.1048 [hep-th]].
%
\\
%
  C.~Sleight and M.~Taronna,
  JHEP {\bf 1801}, 060 (2018)
  [arXiv:1708.08668 [hep-th]].
%
\\
%
  D.~Polyakov,
  Phys.\ Rev.\ D {\bf 93}, no. 4, 045001 (2016)
  [arXiv:1511.04563 [hep-th]].

\bibitem{Metsaev:2012uy}
  R.~R.~Metsaev,
  Phys.\ Lett.\ B {\bf 720}, 237 (2013)
  [arXiv:1205.3131 [hep-th]].








\bibitem{Manvelyan:2010jr}
  R.~Manvelyan, K.~Mkrtchyan and W.~Ruhl,
  Nucl.\ Phys.\ B {\bf 836}, 204 (2010)
  [arXiv:1003.2877 [hep-th]].
%
\\
%
  A.~Sagnotti and M.~Taronna,
  Nucl.\ Phys.\  B {\bf 842}, 299 (2011)
  [arXiv:1006.5242 [hep-th]].
%
\\
%
  R.~Manvelyan, K.~Mkrtchyan and W.~Ruehl,
  Phys.\ Lett.\  B {\bf 696}, 410 (2011)
  [arXiv:1009.1054 [hep-th]].



\bibitem{Vasilev:2011xf}
  M.~A.~Vasiliev,
  Nucl.\ Phys.\ B {\bf 862}, 341 (2012)
  [arXiv:1108.5921 [hep-th]].
%
\\
%
  E.~Joung and M.~Taronna,
  Nucl.\ Phys.\ B {\bf 861}, 145 (2012)
  [arXiv:1110.5918 [hep-th]].
%
\\
  N.~Boulanger, D.~Ponomarev and E.~D.~Skvortsov,
  JHEP {\bf 1305}, 008 (2013)
  [arXiv:1211.6979 [hep-th]].



\bibitem{Basile:2017kaz}
  T.~Basile,
  Universe {\bf 4}, no. 1, 4 (2018)
  [arXiv:1710.10572 [hep-th]].
%
\\
%
  T.~Basile, E.~Joung, S.~Lal and W.~Li,
  JHEP {\bf 1810}, 091 (2018)
  [arXiv:1805.05646 [hep-th]].















\end{thebibliography}
\end{document}